\documentclass[12pt]{article}

\ifx\pdfoutput\undefined
\usepackage[dvips,bookmarks]{hyperref}
\else
\usepackage{hyperref}
\fi
\hypersetup{colorlinks=false,bookmarksopen,bookmarksnumbered,citecolor=blue,
   pdfstartview=FitH}

\usepackage{latexsym}

\usepackage{amssymb,amsfonts,amsmath}
\usepackage{graphicx} 
\usepackage{indentfirst}

 \usepackage{bbm}

\parskip=\medskipamount

\def\a{\alpha}

\def\b{\beta}
\def\c{\chi}
\def\d{\delta}
\def\e{\epsilon}

\def\g{\gamma}
\def\G{\Gamma}

\def\k{\kappa}

\def\p{\pi}
\def\q{\theta}

\def\s{\sigma}

\newcommand{\be}{\begin{equation}}
\newcommand{\ee}{\end{equation}}
\newcommand{\bea}{\begin{eqnarray}}
\newcommand{\eea}{\end{eqnarray}}

\newcommand{\ba}{\begin{array}}
\newcommand{\ea}{\end{array}}

\def\double #1{#1{\hbox{\kern-2pt $#1$}}}

\newcommand{\bsubeq}{\begin{subequations}}
\newcommand{\esubeq}{\end{subequations}}

\begin{document}

\begin{titlepage}

\begin{flushright}
June 2015\\
\end{flushright}
\vspace{5mm}

\begin{center}
{\Large \bf  Coset construction of a D-brane gauge field }
\end{center}

\begin{center}

{\large  
I. N. McArthur\footnote{{ian.mcarthur@uwa.edu.au}}
} \\
\vspace{5mm}

\footnotesize{
{\it School of Physics M013, 
The University of Western Australia\\
35 Stirling Highway, Crawley W.A. 6009 Australia }}  
~\\

\vspace{2mm}

\end{center}
\vspace{5mm}

\begin{abstract}
\baselineskip=14pt
\noindent
$D$-branes have a world-volume $U(1)$ gauge field $A$ whose field strength $F = dA$ gives rise to  a Born-Infeld term in the $D$-brane action. Supersymmetry and kappa symmetry transformations of $A$ are traditionally inferred by the requirement that the Born-Infeld term is consistent with both supersymmetry and kappa symmetry of the $D$-brane action. In this paper, we show that integrability of the assigned supersymmetry transformations leads to a extension of the standard supersymmetry algebra that includes a fermionic central charge. 
 We construct a superspace one-form on an enlarged superspace related by a coset construction to this centrally extended algebra whose supersymmetry and kappa symmetry transformations are derived,  rather than inferred. It is shown that under pullback, these transformations are of the form expected for the $D$-brane $U(1)$ gauge field. We relate these results to manifestly supersymmetric approaches to construction of $D$-brane actions. 
\end{abstract}
\vspace{1cm}

\vfill
\end{titlepage}

\tableofcontents{}
\vspace{1cm}
\bigskip\hrule

\section{Introduction}
\setcounter{equation}{0}

In the Green-Schwarz formulation \cite{Green:1983wt, Green:1983sg, Henneaux:1984mh, DeAzcarraga:1989vh}, $p$-branes are embeddings of a ($p$ + 1)-dimensional bosonic world-volume into a superspace, 
\be
\s ^i \rightarrow \left( x^a (\s), \q^{\a} (\s) \right),
\ee
where $\s^i$ are coordinates on the world-volume, and $(x^a , \q^{\a})$ are superspace coordinates. Here we consider flat  $D$-dimensional ${\cal N} = 1$ superspace. The supersymmetry algebra
\be
\{ Q_{\a}, Q_{\b} \} = - 2 \, (C \G^a)_{\a \b} \, P_a
\ee
is realised via the transformations of  superspace coordinates\footnote{Unless otherwise stated, spinors are Majorana, and $ \bar{\e} = \e^{T} C,$ where $C$ is the charge conjugation matrix. Also, the spacetime dimension must be such that $(C \G^a)_{\a \b}$ is symmetric. }
\bea
\d_{\e} \, x^a &=& i \, (\bar{\e} \G^a \q) \\
\d_{\e} \, \q^{\a} &=&  \e^{\a}.
\label{SS}
\eea
The one-forms 
\be
\p^{a} = dx^a - i (\bar{\q} \G^a d \q), \quad d \q^{\a}
\ee
are invariant under supersymmetry transformations.
$p$-brane actions exist only in spacetime dimensions for which the supersymmetry invariant ($p$+2) form 
\be
h^{(p+2)} = \p^{a_1} \wedge \cdots \wedge \p^{a_p} \, (d \bar{\q} \, \G_{a_1} \cdots \G_{a_p} \, d \q)
\label{h}
\ee
is closed, requiring the gamma matrix identities
\bea
0 & = & ( C\G^{a})_{\a ( \b} \, (C \G_{a} )_{\g \d )} , \quad p = 1 \label{id1} ; \\
0 & = & ( C\G^{a_1})_{(\a \b} \, (C \G_{a_1 \cdots a_p})_{\g \d)},  \quad p > 1.
\eea
Here, $ \G_{a_1 \cdots a_p}$ is the anti-symmetrized product of gamma matrices, and the round brackets on spinor indices denote symmetrisation. The resulting restrictions on $p$ and $D$ give  rise to the ``brane-scan'' \cite{Achucarro:1987nc}.
Closure of $h^{(p+2)}$ implies
 \be
h^{(p+2)} = d \, b^{(p+1)},
\ee
The $p$-brane action is 
\be 
S = S_0 + S_{WZ},
\ee
where the ``kinetic'' term 
\be
S_0 = \int d^{(p+1)} \s \, \sqrt{ det \,G_{ij}}
 \ee
is constructed from the pulled back world-volume metric
\be
G_{ij} = \p_i{}^a  \eta_{ab}   \p_j{}^b, 
\ee
with $\p_i{}^a = \frac{\partial x^a}{\partial \s^i} - i (\bar{\q} \G^a \frac{\partial \q}{\partial \s^i}).$ 
The Wess-Zumino term is given by the integral over the ($p$+1)-dimensional world-volume of the pullback of the superspace form $b^{(p+1)},$
\be
S_{WZ} = \int \s^* b^{(p+1)}.
\label{WZ1}
\ee
 The fact that $b^{(p+1)}$ is  not invariant under supersymmetry transformations, but varies by a total derivative, leads to a central extension to the supersymmetry algebra of the form \cite{de Azcarraga:1989gm}
\be
\{ Q_{\a}, Q_{\b} \} = - 2 \, (C \G^a)_{\a \b} \, P_a + (C \G_{a_1 \cdots a_p})_{\a \b} \, Z^{a_1 \cdots a_p} ,
\ee
where $Z^{a_1 \cdots a_p} $ are bosonic central charges.

It is possible to construct Wess-Zumino Lagrangians that are manifestly invariant under supersymmetry transformations as forms on enlarged superspaces related by the standard coset construction of superspaces to centrally extended supersymmetry algebras \cite{Siegel:1994xr, Bergshoeff:1995hm, Reimers:2005jf, Reimers:2005cp}.  We will come back to this point later.

The $p$-brane action admits a local fermionic symmetry called $\k$-symmetry. For the action constructed in standard superspace (as opposed to the enlarged superspaces associated with central extensions of the supersymmetry algebra) with the Wess-Zumino term (\ref{WZ1}), the $\k$-symmetry transformations take the form
\bea
\d_{\k} \q^{\a} &=& \k^{\a} \\
\d_{\k} x^a &=& - i (\bar{\k} \G^a \q),
\eea
where $\k^{\a}$ involves a projection operator onto half of the fermionic degress of freedom. This allows these degrees of freedom to be gauged away, achieving a matching of numbers of  physical bosonic and fermionic degrees of freedom on the world-volume, as required for world-volume supersymmetry \cite{de Azcarraga:1982dw, de Azcarraga:1982xk, Siegel:1983hh}. In terms of the coset construction of standard superspace, $\k$-symmetry has a natural interpretation as a right group action, thus being equivalent to an enlargement of the isotropy group \cite{McArthur:1999dy, deAzcarraga:2004df}.

As originally formulated, $D$-branes\footnote{The ``$D$" in this context refers to ``Dirichlet,'' a choice of boundary conditions for open superstrings, and has nothing to do with the spacetime dimension.} include a world-volume  $U(1)$ gauge field $A$ whose field strength $F = d A$ gives rise to a Born-Infeld term in the action \cite{Townsend:1995af, Aganagic:1996pe, Aganagic:1996nn, Cederwall:1996pv}; specifically,
\be
S =  \int d^{(p+1)} \s \, \sqrt{det \,(G_{ij}+ {\cal F}_{ij})} + S_{WZ}.
\ee
As with $p$-branes,  $G_{ij}$ is the pullback to the world-volume of the spacetime metric.  $ {\cal F} $ is the world-volume two-form ${\cal F} = F -  \s^*b^{(2)}, $ where $ d b^{(2)} = h^{(3)} $ is a supersymmetry invariant closed superspace three-form given by (\ref{h}) with $p = 1$. The Wess-Zumino term is again of the form
\be
S_{WZ} = \int \s^* b^{(p+1)},
\ee
with $h^{(p+2)} = db^{(p+1)}$ also supersymmetry invariant and closed as in (\ref{h}). Explicit details of the construction depend on the nature of the superspace into which the $D$-brane is embedded; however, a worldvolume $U(1)$ gauge field $A$ is  required in all cases.  $D$-branes also possess a $\k$-symmetry, which again allows half of the fermionic degrees of freedom to be gauged away.

In this formulation of $D$-brane actions, the supersymmetry transformations and $\k$-symmetry transformations of  the world-volume one-form $A$ are {\it inferred} on the basis of the requirement that ${\cal F}= F - \s^* b^{(2)} $
is invariant under supersymmetry transformations and that ${\cal F}$ has a $\k$-symmetry transformation of the form
\be
\d_{\k} {\cal F} \sim (\bar{\k} \G_a d \q) \wedge \p^a
\label{dcalF}
\ee
(appropriately pulled back). 
The former ensures the $D$-brane action is invariant under spacetime supersymmetry transformations, and the latter is needed to allow the action including the Born-Infeld term to be made $\k$-symmetric in order to ensure equal numbers of world-volume bosonic and fermionic degrees of freedom - as required for world-volume supersymmetry. Inclusion of the bosonic degrees of freedom associated with the world-volume guage field $A$ gives rise to a revised ``brane scan" for $D$-branes \cite{Duff:1992hu}. It is also required that supersymmetry transformations and $\k$-symmetry transformations commute \cite{Cederwall:1996pv}. The precise details of the construction depend on the nature of the superspace into which the $D$-brane is embedded - see \cite{Townsend:1995af, Aganagic:1996pe, Aganagic:1996nn, Cederwall:1996pv} for the case of 10-dimensional flat superspace. However,  these constructions require an {\it assignment} of spacetime supersymmetry and $\k$-symmetry transformations to the world-volume gauge field $A$ rather than a derivation of these transformations. Assigning spacetime transformations to a purely world-volume degree of freedom seems somewhat unnatural.

Subsequent to the original formulation of $D$-brane actions, an alternative approach based on enlarged superspaces associated with central extensions of the standard supersymmetry algebra was developed \cite{Sakaguchi:1998kk, Sakaguchi:1998sy, Chryssomalakos:1999xd, Sakaguchi:1999fm, Hatsuda:2000vb, deAzcarraga:2001fi, Reimers:JPhysA, Reimers:2005fp}. These constructions relied only on pull-backs of superspace geometry, and in particular did not require the existence of a purely world-volume gauge field $A.$

The outcomes of this paper are two-fold. Starting with the original formulation of $D$-brane actions involving a world-volume gauge field,  we provide a natural motivation for a construction of the Born-Infeld term based on an enlarged superspace. Specifically, we show that integrability of  transformations conventionally assigned to the world-volume gauge field to ensure spacetime supersymmetry leads to the requirement for a realisation of a fermionic central extension of the supersymmetry algebra  \cite{Green:1989nn}  first introduced by Green in the context of superstring theory. We are led to introduce an enlarged superspace with coordinates $(x^a, \q^{\a}, \c_{\a})$ related by a coset construction to the Green algebra.   In the case of a flat supersymmetric ${\cal N} = 1$ superspace for which the gamma matrix identity (\ref{id1}) is true, we show this allows construction of a superspace one-form which upon pullback yields a world-volume one-form with the spacetime supersymmetry transformations imposed somewhat arbitrarily in the original construction of $D$-brane actions.
 We relate these results to later approaches to constructing $D$-brane actions which take as their starting point an enlarged superspace as a means to ensure manifest spacetime supersymmetry of Born-Infeld terms.

We also provide an explicit construction of kappa symmetry transformations of the world-volume gauge field (considered as a pullback of a one-form defined in an enlarged superspace) in terms of a right group action, extending the interpretation provided in \cite{McArthur:1999dy, deAzcarraga:2004df} for $\k$-symmetry transformations for $p$-brane actions. Again, we can relate this to previous work on manifestly supersymmetry invariant formulations of $D$-brane Lagrangians based on enlarged superspace.

\section{Integrability of the supersymmetry transformation of the world-volume gauge field}
\setcounter{equation}{0}

We begin with the supersymmetry invariant and closed three-form
\be
 h^{(3)} = \p^a \wedge (d \bar{\q} \G_a d \q) 
 \ee
in flat ${\cal N} =1$ superspace -- closure relies on the gamma matrix identity (\ref{id1}). In fact, using this identity,
\be
 h^{(3)} = dx^a \wedge (d \bar{\q} \G_a  d \q).
 \ee
Since $h^{(3)}$ is closed and invariant under supersymmetry transformations, locally
\be
h^{(3)} = d b^{(2)}
\ee
with 
\be
\d_{\e} b^{(2)} = d a^{(1)}(\e),
\ee
where $\d_{\e}$ denotes a supersymmetry transformation with parameter $\e^{\a}$. We introduce a one-form $A$ whose supersymmetry transformation is determined by the requirement that  $ { \cal F} = dA - b^{(2)} $ is invariant. Then
\be
\d_{\e} A = a^{(1)} (\e).
\ee

In the case $h^{(3)} = dx^a \wedge (d \bar{\q} \G_a  d \q),$
one can see by inspection that a candidate for $b^{(2)}$ is
\be
b^{(2)} = \a \,x^a (d \bar{\q} \G_a d \q) - (1 - \a) \, dx^a \wedge (\bar{\q} \G_a  d \q),
\label{b2}
\ee
where $\a$ is a real parameter. Varying $\a$ changes $b^{(2)} $ by an exterior derivative, so it is an ``integration constant'' in solving  $h^{(3)} = d b^{(2)}.$  Conventionally $\a$ is set to zero in the literature. It is straightforward to show that 
$ \d_{\e} b^{(2)} = d a^{(1)} (\e) $ with 
\be
a^{(1)} (\e) = (1- \a) \b \, dx^a (\bar{\e} \G_a \q) - \frac{i}{3} (1 - 3 \a) \, (\bar{\e} \G^a \q) (\bar{\q} \G_a d \q) - (1 - \a)(1 - \b) \, x^a (\bar{\e} \G_a d \q),
\ee
with $\b$ again an ``integration constant'' appearing in the cohomology (conventionally set to 1); the only nontrivial step is use of the gamma matrix identity (\ref{id1}) to show 
\be
(\bar{\e} \G^a \q) (d \bar{\q} \G_a d \q) = 2 (\bar{\e} \G^a d  \q) (\bar{\q} \G_a d \q).
\label{dtdt}
\ee
So we arrive at the proposed supersymmetry transformation for the one-form $A,$
\be
\d_{\e} A = (1- \a) \b \, dx^a (\bar{\e} \G_a \q) - \frac{i}{3} (1 - 3 \a) \, (\bar{\e} \G^a \q) (\bar{\q} \G_a d \q) - (1 - \a)(1 - \b) \, x^a (\bar{\e} \G_a d \q).
\label{deA}
\ee

We can check the integrability of this transformation by computing the commutator of two supersymmetry transformations:
\be 
(\d_{\e_2} \d_{\e_1} - \d_{\e_1} \d_{\e_2} ) A = 2 (\bar{\e_1} \G_a \e_2 )\, \left( (1 - \a) \b \, dx^a + i \a \, (\bar{\q} \G^a d \q) \right).
\label{ddA}
\ee
The only nontrivial step again involves the identity (\ref{id1}) to show
\be
(\bar{\e_1} \G^a  \q) (\bar{\e_2} \G_a  d \q) - (\bar{\e_2} \G^a \q) (\bar{\e_1} \G_a d \q) = (\bar{\e_1} \G^a \e_2 )(\bar{\q} \G_a d \q).
\
\ee
If we use $\d_{\e} =  \e^{\a} Q_{\a} ,$ and require consistency of (\ref{ddA}) with the anticommutator $\{ Q_{\a}, Q_{\b} \} = -2 (C \G^a)_{\a \b} P_a,$
then we infer that as an operator relation in ``$(x, \q, A)$ space"\footnote{ Compatibility of $(\d_{\e_2} \d_{\e_1} - \d_{\e_1} \d_{\e_2} ) x^a = 2 i (\bar{\e_1} \G^a \e_2 ) $ with the algebra requires $P_a x^b = - i \d_a{}^b$},
\be 
P_a A = - (1 - \a) \b dx_a - i \a (\bar{\q} \G_a d \q).
\label{PaA}
\ee
We can then compute that
\be
[ P_a , \d_{\e} ] A =  i (\bar{\e} \G_a d \q ).
\ee
Note that the result turns out to be independent of the choices of the ``integration constants'' $\a$ and $\b$ introduced in solving the cohomology equations, showing the result is inherent to the initial cohomological problem.
With $\d_{\e} = \e^{\a} Q_{\a},$ this means 
\be
[ P_a ,Q_{\a} ] A = i (C \G_a  d\q )_{\a},
\label{PQA}
\ee
which is not consistent with the the standard supersymmetry algebra which has $[ P_a ,Q_{a} ] = 0.$ Instead, we have a realisation of a central extension of the standard supersymmetry algebra. In particular, since $[ P_a ,Q_{a} ] $ is fermionic, we require a fermionic central charge.\footnote{ In inferring the form of (\ref{PaA}) of $P_a A,$ we assumed no central extension to  $\{ Q_{\a}, Q_{\b} \} = -2 (C \G^a)_{\a \b} P_a.$ By redefining $P_a A $ (equivalent to introducing a central extension to the anticommutator $\{ Q_{\a}, Q_{\b} \} $), it is possible to ensure $[ P_a ,Q_{a} ] A = 0.$ }

It is interesting to note that the expression for $\d_{\e} b^{(2)}$ computed from (\ref{b2}) using the transformation rules (\ref{SS}) is integrable, in that it yields a representation of the standard supersymmetry algebra without central extension. The central extension enters only when we try to solve $ \d_{\e} b^{(2)} = d \, \d_{\e} A, $ with $b^{(2)}$ determined by $ 0 = \d_{\e} h^{(3)} = d \, \d_{\e} b^{(2)}.$ 

\section{The ``Green'' algebra and coset construction}
\setcounter{equation}{0}
A central extension of the supersymmetry algebra in the context of superstring theory with a fermionic central charge $Z^{\a}$ was considered by Green \cite{Green:1989nn}:
\bea
\, \{ Q_{\a}, Q_{\b} \} &=& -2 (C \G^a)_{\a \b} P_a \\
\, [ P_a,  Q_{\a}] &=&   i (C \G_a)_{\a \b} \, Z^{\b}.
\label{GA}
\eea
The Jacobi identity $[ Q_{(\a }, \{ Q_{\b}, Q_{\g)} \} ] = 0 $ (with round brackets denoting symmetrisation of spinor indices) is satisfied as a result of the gamma matrix identity (\ref{id1}).
In order for  the result (\ref{PQA}) to reflect this centrally extended algebra, we require that 
\be
Z^{\a} A =  d \q^{\a}.
\label{ZaA}
\ee
We will assume that the superspace coordinates $x^a$ and $\q^{\a}$ are inert under the charge $Z^{\a}.$

There is a natural way to achieve this result. If $ \c_{\a}$ is a Goldstone boson for breaking of the symmetry generated by $Z^{\a},$ then 
\be
Z^{\a}  \c_{\b} = \d^{\a}{}_{\b}
\label{Zl}
\ee
(meaning that $\c_{\a}$ shifts by a constant spinor under transformations generated by $Z^{\a}$). In this case,
\be A =  (\bar{\c} d \q) + \cdots 
\ee
would fulfil the condition (\ref{ZaA}) (the dots represent potential contributions to the one-form involving $x, \q, dx $ and $d \q,$ which are assumed inert under the action of $Z^{\a}$).

This suggests seeking a nonlinear realisation of the Green algebra using standard techniques based on a coset construction. This corresponding enlarged superspace has been considered previously  \cite{Siegel:1994xr, Bergshoeff:1995hm}. We introduce the group element
\be 
g(x, \q, \c) = e^{i(x^a P_a + \q^a Q_{\a} -  \c_{\a} Z^{\a} )}.
\label{extcoset}
\ee
A supersymmetry transformation with parameter $\e$ is then achieved by left action of $e^{i \e^{\a} Q_{\a}}$ on $g(x, \q, \c).$ Using the Baker-Campbell-Hausdorff formula, we find that in infinitesimal form,
\bea
\, \d_{\e} \q^{\a} & = &  \e^{\a}  \\
\, \d_{\e} x^a & = &  i (\bar{\e} \G^a \q) \\
\, \d_{\e}  \c_{\a} & = &  - \frac12 x^a (\bar{\e} \G_a )_{\a} + \frac{i}{6} (\bar{\e} \G^a \q ) (\bar{\q} \G_a)_{\a} .
\label{de}
\eea
It is then straightforward to check that if we set 
\be
A = \c_{\a} d \q^{\a} - i (1 - \a ) \b \,  x^a d x_a + (\a - \frac12) \, x^a ( \bar{\q} \G_a d \q),
\label{A}
\ee
we reproduce the supersymmetry transformation law (\ref{deA}), which, we recall, was initially inferred based on the requirement that $ {\cal F} = dA - b^{(2)}$ be invariant under supersymmetry transformations.

The supersymmetry transformations (\ref{de}) can be reproduced by the action of a differential operator $\e^{\a} {\cal Q}_{\a}$ on an enlarged superspace with coordinates $(x^a, \q^{\a}, \c_{\a}) $, with
\be
{\cal Q}_{\a} = \frac{\partial}{\partial \q^{\a} } + i (C \G^a \q)_{\a} \frac{\partial}{\partial x^a } - \frac12 x^a (C \G_a)_{\a \b} \frac{\partial}{\partial \c_{\b} } + \frac{i}{6} (C \G^a \q)_{\a} (\bar{\q} \G_a)_{\b} \frac{\partial}{\partial \c_{\b} }.
\label{Q}
\ee
Then the anticommutator of two supersymmetry generators takes the form (using the gamma matrix identity (\ref{id1}) )
\be
\{ {\cal Q}_{\a}, {\cal Q}_{\b} \} = -2 (C \G^a)_{\a \b} {\cal P}_a,
\ee
with \be
{\cal P}_a = - i \frac{\partial }{\partial x^a} - \frac{i}{2} (\bar{\q} \G_a)_{\a}   \frac{\partial}{\partial \c_{\a} }.
\label{cP}
\ee
The reason for the presence of a $\c$ derivative in the spacetime translation operator is that $\c_{\a}$ transforms nontrivially due to the central extension in the supersymmetry algebra. If we consider
\be
e^{i a^a P_a } \, g(x, \q, \c),
\ee
we find spacetime transformations
\bea
\d_a x^a &=&  a^a \\
\d_a \q^{\a} &=& 0 \\
\d_a \c_{\a} &=& \frac12 a^a (\bar{\q} \G_a)_{\a},
\eea
which are reproduced by the action of the differential operator $\d_a = i a^a {\cal P}_a$ on the enlarged $(x^a, \q^{\a}, \c_{\a})$ superspace.
Indeed, if we use expression (\ref{A}) for the one-form $A,$ we find
\be
{\cal P}_a A = - (1 - \a) \b dx^a - i \a (\bar{\q} \G_a d \q),
\ee
which reproduces the earlier result (\ref{PaA}).

Finally, using the differential operators $ {\cal P}_a$ and $ {\cal Q}_{\a}$ above, we compute that
\be
[ {\cal P}_a, {\cal Q}_{\a} ] = i (C \G_a)_{\a \b} \frac{\partial}{\partial \c_{\b}},
\ee
which is consistent with the Green algebra (\ref{GA}) providing the central charge $Z^{\a}$ is represented by the action of the differential operator
\be
{\cal Z}^{\a} = \frac{\partial}{\partial \c_{\a}}
\ee
on the enlarged superspace. This is consistent with the earlier speculation 
(\ref{Zl}).

\section{Deriving the $\k$-symmetry transformation of $A$}
\setcounter{equation}{0}

On the basis of the construction (\ref{A}) for the one-form $A,$ we can go further - we can determine the $\k$-symmetry transformation of $A$ from first principles. The $\k$-symmetry transformations of $x^a$ and $\q^{\a}$ look like supersymmetry transformations, except that the sign of the transformation of $x^a$ is reversed \cite{Green:1987sp}. In  \cite{McArthur:1999dy}, it was shown that this has a natural interpretation in terms of the right action of the super-Poincar\'e group on superspace coset representatives. In particular, since supersymmetry transformations relate to the left action of the group, and left and right actions commute, this automatically ensures supersymmetry transformations and $\k$-symmetry transformations commute. Applying the same philosophy here, we postulate that the  $\k$-symmetry transformations are generated by a right action
\be
g( x , \q, \c) \rightarrow g (x, \q, \c) \, e^{i \k^{\a} Q_{\a}}.
\ee
Using the Baker-Campbell-Hausdorff formula, this yields the following infinitesimal $\k$-symmetry transformations:
\bea
\d_{\k} \theta^{\a} & = & \k^{\a} \\
\d_{\k} x^a &=& -i \, (\bar{\k} \G^a \q) \\
\d_{\k}  \c_{\a} & = &   \frac12 x^a (\bar{\k} \G_a )_{\a} + \frac{i}{6}  (\bar{\k} \G^a \q ) (\bar{\q} \G_a)_{\a} .
\eea
Applying these to the one-form $A$ as defined in (\ref{A}), we find
\be
\d_{\k} A =  ( \a - \b + \a \b ) x^a (\bar{\k} \G_a d \q) - (1 - \a) \b \, dx^a (\bar{\k} \G_a \q) + i (\frac23  -  \a) (\bar{\k} \G^a  \q) (\bar{\q} \G_a d \q).
\ee
From this it follows that for $F = dA,$
\be
\d_{\k} F = -  \a (\bar{\k} \G_a d \q) \wedge dx^a + i (2 - 3 \a) \, (\bar{\k} \G_a d \q) \wedge (\bar{\q} \G^a d \q),
\ee
involving use of the identity (\ref{dtdt}).
On the other hand, applying the $\k$-symmetry transformations to (\ref{b2}),
 we find
 \be
 \d_{\k} b^{(2)} = (1 - \a) (\bar{\k} \G_a d \q) \wedge dx^a + i (1 - 3 \a) (\bar{\k} \G_a d \q) \wedge (\bar{\q} \G^a d \q).
 \ee
 Combining these results, we find that for ${\cal F} = F - b^{(2)},$
 \be
 \d_{\k} {\cal F } = - (\bar{\k} \G_a d \q) \wedge (d x^a - i (\bar{\q} \G^a d \q) ) = - (\bar{\k} \G_a d \q) \wedge \pi^{a},
 \ee
 exactly as required in (\ref{dcalF})  in order to engineer $\k$-symmetry of the $D$-brane action.

\section{Relationship to manifestly spacetime supersymmetric constructions}
\setcounter{equation}{0}
In this section, we show how the results obtained above based on the original formulation of $D$-brane actions  \cite{Townsend:1995af, Aganagic:1996pe, Aganagic:1996nn, Cederwall:1996pv} are related to a subsequent approaches based on embedding of the world-volume in an enlarged superspace which does not require the introduction of an explicit world-volume gauge degree of freedom and in which spacetime supersymmetry is manifest \cite{Sakaguchi:1998kk, Sakaguchi:1998sy, Chryssomalakos:1999xd, Sakaguchi:1999fm, Hatsuda:2000vb, deAzcarraga:2001fi, Reimers:JPhysA, deAzcarraga:2004df, Reimers:2005fp}. The latter approaches extended earlier work on manifestly supersymmetric Wess-Zumino terms for $p$-branes \cite{Siegel:1994xr, Bergshoeff:1995hm, Reimers:2005jf, Reimers:2005cp}. 

Using the coset parameterisation (\ref{extcoset}) of the enlarged superspace related to the Green algebra,
\be
g(x, \q, \c) = e^{i(x^a P_a + \q^a Q_{\a} -  \c_{\a} Z^{\a} )},
\ee
the corresponding Maurer-Cartan form is
\be
g(x, \q, \c)^{-1} d g(x, \q, \c) = i \left( E^a P_a + E^{\a} Q_{\a} + \tilde{E}_{\a} Z^{\a} \right),
\label{MC}
\ee
with
\bea
E^a &=& d x^a - i (\bar{\theta} \G^a d \theta ) \\
E^{\a} &=&  d \theta^{\a} \\
\tilde{E}_{\a}  &=& d  \c_{\a} - \frac12 x^a  (d \bar{\theta} \G_a)_{ \a} + \frac12   dx^a (\bar{\theta} \G_a)_{ \a} - \frac{i}{3} ( \bar{\theta} \G^a d \theta) (\bar{\theta} \G_a) _{\a}.
\eea
It is easy to show that the Born-Infeld contribution to the $D$-brane action ${\cal F} = dA - b^{(2)},$ with $A$ and $b^{(2)}$ defined in (\ref{A}) and (\ref{b2}),  can be expressed in terms of the Maurer-Cartan forms for the extended Green superspace as 
\be 
{\cal F } = \tilde{E}_{\a} E^{\a}.
\label{extBI}
\ee
This is manifestly spacetime supersymmetry invariant as the Maurer-Cartan form (\ref{MC}) is invariant under the global left action $  g(x, \q, \c) \rightarrow e^{i \bar{\epsilon} Q} \, g(x, \q, \c).$ 

A manifestly supersymmetric construction of the Born-Infeld contribution to a $D$-brane was  first by provided Sakaguchi \cite{Sakaguchi:1998kk}, based on a dimensional reduction of the ``$M$-algebra'' of Sezgin \cite{Sezgin:1996cj}. The algebra considered by Sakaguchi is much larger than the Green algebra. In this paper, we provide a rationale for the appearance of an enlarged algebra - the integrability of the spacetime supersymmetry transformations (\ref{deA}) assigned to $A$ - and realise this with with a minimal extension of the standard supersymmetry algebra, namely  the Green algebra.

Similarly, $\k$-symmetry of the Born-Infeld term  has previously been considered in the context of a local right action on an enlarged superspace, \cite{Reimers:2005fp}, building on  this interpretation of $\k$-symmetry for $p$-branes in \cite{McArthur:1999dy, deAzcarraga:2004df}. 
Again, this paper provides a bridge between the original formulations of $\k$-symmetry for the gauge field in the $D$-brane action and this later work via a rationale for consideration of an enlarged superspace related to the Green algebra. Further, we realise the $\k$-symmetry on a minimal enlargement of superspace based on the Green algebra, rather than more general algebras considered in \cite{Reimers:2005fp}.

\section{Conclusion}
\setcounter{equation}{0}
In this paper, we have analysed the spacetime supersymmetry transformations somewhat arbitrarily assigned to the world-volume gauge field in the original formulation of  $D$-brane actions. We showed that integrability of these transformations automatically leads to a fermionic central extension of the standard supersymmetry algebra, the Green algebra. This provides a rationale for the consideration of  an enlarged coset superspace based on the Green algebra. We have constructed a one-form in this enlarged superspace, which, when pulled back to the world-volume,  has the supersymmetry and $\k$-symmetry transformations  required to construct the Born-Infeld terms in $D$-brane actions. We have also shown how this ``bottom up'' construction of forms in an expanded superspace relates to earlier approaches to $D$-brane actions based on manifestly spacetime supersymmetric Wess--Zumino terms, which also involve enlarged superspaces, though based on algebras larger than the Green algebra.

In \cite{Freed:1999vc}, Freed and Witten show that the world-volume gauge field incorporated in the conventional construction of $D$-brane actions is not a normal $U(1)$ gauge field. Rather, it defines a ``Spin$^c$ structure'' on the world-volume, and is also intimately involved in cancellation of certain anomalies. It would be of interest to see if the superspace one-form constructed in this paper is still able to fulfil these requirements by pullback.

Another crucial issue is that a $D$-brane should admit a world-volume supersymmetry (as opposed to that of the superspace into which the world-volume is embedded). This requires equal numbers of world-volume bosonic and fermionic degrees of freedom (giving rise a ``$D$-brane scan" \cite{Duff:1992hu} determining allowed relationships between world-volume and spacetime dimensions for various extended supersymmetries). 
In the conventional approach to construction of $D$-brane actions, the gauge field lives on the $(p+1)$-dimensional world-volume and appears in the action only via the corresponding curvature, giving rise to $p-1$ world-volume bosonic degrees of freedom (complementing the $D - p - 1$ bosonic Goldstone degrees of freedom arising from the embedding of a $(p+1)$-dimensional world-volume into a $D$-dimensional spacetime). However, this world-volume gauge field is attributed the spacetime supersymmetry transformation (\ref{deA}) and $\k$-symmetry transformation (\ref{dcalF}), which is somewhat artificial for a world-volume degree of freedom. In this paper, we have provided a construction of a superspace one-form which naturally yields the transformations (\ref{deA}) and  (\ref{dcalF}), and which upon pullback furnishes a one-form degree of freedom on the world-volume which appears in the action only via the corresponding curvature. This one-form lives in an enlarged superspace, and there are  gauge symmetries associated with local right actions on the coset representatives of points in this superspace ($\k$-symmetry is one such symmetry). It remains to be investigate whether the gauge symmetry associated with the additional fermionic generator in the Green algebra givea rise to a balance between the worldvolume bosonic and fermionic  degress of freedom required for worldvolume supersymmetry. Note that  this issue has been touched upon in \cite{deAzcarraga:2001fi}, though in  superspaces  larger than that based on the Green algebra.

We have only considered $D$-branes propagating in a flat superspace. A central extension of the standard flat superspace algebra, the Green algebra \cite{Green:1989nn}, has been used  to provide a coset construction for the supersymmetry and $\k$-symmetry transformations of the $U(1)$ gauge field  necessary for construction of $D$-brane actions. It is natural to enquire how this construction might be extended to  $D$-branes propagating in curved superspaces. Detailed analysis has been undertaken for $D$-branes propagating in curved superspaces constructed as cosets for superconformal extensions of the flat superspace algebra \cite{Metsaev:1998it,Metsaev:1998hf,Metsaev:2000bj,Metsaev:2002xc}; and indeed, the formulation of $\k$-symmetry as a right group action in \cite{McArthur:1999dy} was for general coset superspaces. It would be of interest to consider whether ``Green-type'' central extensions of these superconformal algebras exist and whether they can be used to construct from first principles the supersymmetry and $\k$-symmetry transformations of the $U(1)$ gauge field in the action for  $D$-branes propagating in the corresponding curved coset superspace.

\noindent
{\bf Acknowledgements:}\\
I wish to thank Daniel Reimers for making me aware of the Green algebra a long time ago.

\begin{footnotesize}

\end{footnotesize}

\end{document}